\documentclass[print,showpacs,twocolumn,superscriptaddress]{revtex4}
\usepackage{amsmath}
\usepackage{graphicx}

\begin{document}

\title{Quantum information processing with a single photon by input-output
process regarding low-Q cavities}
\author{Jun-Hong An}
\email{phyaj@nus.edu.sg}
\affiliation{Centre for Quantum Technologies and Department of Physics, National
University of Singapore, 3 Science Drive 2, Singapore 117543, Singapore}
\affiliation{Department of Modern Physics, Lanzhou University, Lanzhou 730000, China}
\author{M. Feng}
\email{mangfeng1968@yahoo.com} \affiliation{State Key Laboratory of
Magnetic Resonance and Atomic and Molecular Physics, Wuhan Institute
of Physics and Mathematics, Chinese Academy of Sciences, Wuhan
430071, China}
\author{C. H. Oh}
\email{phyohch@nus.edu.sg}
\affiliation{Centre for Quantum Technologies and Department of Physics, National
University of Singapore, 3 Science Drive 2, Singapore 117543, Singapore}

\begin{abstract}
Both cavity QED and photons are promising candidates for quantum
information processing. We consider a combination of both
candidates with a single photon going through spatially separate
cavities to entangle the atomic qubits, based on the input-output
process of the cavities. We present a general expression for the
input-output process regarding the low-Q cavity confining a single
atom, which works in a wide range of parameters. Focusing on
low-Q cavity case, we propose some schemes for quantum information
processing with Faraday rotation using single photons, which is much
different from the high-Q cavity and strong coupling cases.
\end{abstract}

\maketitle

\section{Introduction}

Over past decades, cavity quantum electrodynamics (QED) has become
an important platform to demonstrate quantum characteristics of atoms
and photons \cite{haro1}. Some remarkable experiments, such as
strong coupling of the trapped atoms with cavity \cite{blockade,
interf}, conditional logic gating
\cite{haro2}, efficient generation of single photons \cite%
{single,singlephoton} and so on have been performed successfully.

With atoms strongly interacting with the local cavity mode as quantum nodes
and the photons flying between different nodes as quantum bus, we may set up
a quantum network, which has been considered as a promising way to scaling of the
qubit system for large-scale quantum information processing. Many
theoretical \cite{cirac,Duan03,pan1,Duan04,Feng,goto} and experimental works \cite%
{interf, beu} have been done in this direction and some advances have been
achieved over recent years. Among the theoretical work mentioned above, the
single photon experiencing input-output process of a cavity \cite%
{Duan04,Feng} is of particular interests, in which the atomic qubits could
be entangled by a single moving photon and the successful implementation is
monitored by a click of the detector.

The present paper will focus on the input-output process regarding optical
cavities. Motivated by a very recent experiment with microtoroidal resonator
(MTR) \cite{Dayan08}, we intend to carry out some quantum information
processing (QIP) tasks by means of the input-output process relevant to
optical cavities with low-Q factors. We have noticed that the proposals \cite%
{Duan04,Feng} required the cavities to be with high quality and with
strong coupling to the confined atoms, otherwise the schemes would
not work well or would be pretty inefficient. Some simulation has been done to check
how well the cavity input-output process works \cite {goto}. It was shown that
the gating time should be much shorter than the decay time of the cavity if we expect
to have the gating with high success probability \cite {goto}. Evidently, it is not an easy experimental
task to meet those requirements because the
efficient output of photons, to some extent, implies the larger
cavity decay rate, i.e., the cavity with a relatively lower Q factor. In this
sense, the recent achievement of the MTR gives us hopes to solve the
problem. Although it is still of large decay rate (i.e., called
`bad' cavity in \cite{Dayan08}) and moderate coupling to the atom,
the MTR, with individual photons input and output through a
microresonator, has explicitly shown the effect of the photon
blockade. So this MRT seems a promising candidate setup to
be meeting the requirements in those QIP schemes \cite{Duan04,Feng}.

Our present work will, however, show the possibility of
accomplishing some interesting QIP tasks with the currently achieved
MTR. The key step is to design a scheme for entangling two atoms
confined respectively in two spatially separate cavities with low-Q
factors. So our work is actually not only relevant to the MTR, but
also related to any single-sided optical cavities with one wall
perfectly reflective but the other partially reflective
\cite{kimble95}. As a result, the MTR and the single-sided optical
cavity will be mentioned alternately in what follows. We will first
present an analytical expression for the reflection rate of the
input-output process, which works for a wide range of parametric
variation, from weak to strong coupling regimes, and in the presence
or absence of the confined atom. Then we will try to use a single
photon to entangle two atoms confined respectively in two spatially
separate cavities, based on which further QIP tasks could be carried
out. We argue that QIP with single photons is an efficient way in
the low-Q cavity situation, which is considerably different from the
high-Q cavity and strong-coupling cases.

Different from the previous schemes \cite{Duan04,Feng} with photonic
polarization unchanged or changed by a phase $\pi$, the large cavity
decay and moderate coupling in our case lead to a certain angle
rotation of the photonic polarization after the input-output
process, which is called Faraday rotation. The Faraday rotation was
originally studied in a resonant medium with Zeeman level
splitting under the radiation of linearly polarized light
\cite{Countens} and has been recently observed experimentally in
cold atom and quantum dot systems
\cite{Labeyrie01,Julsgaard01,Li06,Imamoglu07}. The key point for the
Faraday rotation is the bi-refringent propagation of the light
through the medium. By using the values from \cite{Dayan08}, we will
show the desired Faraday rotation for our purpose can be obtained
in a two-mode cavity by using single photons with suitable
frequencies.

The paper is organized as follows. In Sec. II, we introduce a
general input-output relation for a single photon with respect to a
low-Q optical cavity confining a single atom under Jaynes-Cummings
model. We show in Sec. III the Faraday rotation in a two-mode cavity
holding a three-level atom by $\Lambda$-type configuration. Based on
the Faraday rotation, we may entangle remote atoms by a single
photon, as shown in Sec. IV. Sec. V is devoted to the application of
the generated entanglement. Finally, we end with some discussion and
a summary in Sec. VI.

\section{Input-output relation under Jaynes-Cummings model}

In this section, we present the basic input-output relation for a
cavity coherently interacting with a trapped two-level atom. Under
the Jaynes-Cummings model, we have the following Hamiltonian,
\begin{equation}
H=\frac{\hbar \omega _{0}}{2}\sigma _{z}+\hbar \omega _{\mathrm{c}}a^{\dag
}a+i\hbar g(a\sigma _{+}-a^{\dag }\sigma _{-}),  \label{H}
\end{equation}%
where $a$ and $a^{\dag }$ are the annihilation and creation operators of the
cavity field with frequency $\omega _{c}$, respectively; $\sigma _{z}$, $%
\sigma _{+}$, and $\allowbreak \sigma _{-}$ are, respectively, inversion,
raising, and lowering operators of the two-level atom with frequency
difference $\omega _{0}$ between the two levels.

Consider a single-photon pulse with frequency $\omega _{\mathrm{p}}$
input in an optical cavity. The pulse can be expressed by $|\Psi _{\mathrm{p}%
}\rangle =\int_{0}^{T}f(t)a_{\mathrm{in}}^{\dag }(t)dt|\mathrm{vac}\rangle $
\cite{Duan04}, where $f(t)$ is the normalized pulse shape as a function of $%
t $, $T$ is the pulse duration, $a_{\mathrm{in}}^{\dag }(t)$, a
one-dimensional field operator, is the cavity input operator satisfying the
commutation relation $[a_{\mathrm{in}}(t),a_{\mathrm{in}}^{\dag }(t^{\prime
})]=\delta (t-t^{\prime })$, and we denote the vacuum of all the optical
modes by $|\mathrm{vac}\rangle $. In the rotating frame with respect to the
frequency of the input pulse, the quantum Langevin equation of the cavity
mode $a$ driven by the corresponding cavity input operator $a_{\mathrm{in}%
}(t)$ is \cite{Walls94}
\begin{equation}
\dot{a}(t)=-[i(\omega _{c}-\omega _{\mathrm{p}})+\frac{\kappa }{2}%
]a(t)-g\sigma _{-}(t)-\sqrt{\kappa }a_{\mathrm{in}}(t),  \label{eom}
\end{equation}%
where $\kappa $ is the cavity damping rate. Moreover, the atomic lowering
operator also obeys a similar equation and, in the rotating frame of the
frequency $\omega _{\mathrm{p}}$, we have,
\begin{equation}
\dot{\sigma}_{-}(t)=-[i(\omega _{0}-\omega _{\mathrm{p}})+\frac{\gamma }{2}%
]\sigma _{-}(t)-g\sigma _{z}(t)a(t)+\sqrt{\gamma }\sigma _{z}(t)b_{\mathrm{in%
}}(t),  \label{eom2}
\end{equation}%
where $b_{\mathrm{in}}(t)$, with the commutation relation $[b_{\mathrm{in}%
}(t),b_{\mathrm{in}}^{\dag }(t^{\prime })]=\delta (t-t^{\prime })$,
is the vacuum input field felt by the two-level atom, $\gamma $ is
the decay rate of the two-level atom. The input and output fields of
the cavity are related by the intracavity field as \cite {Walls94}
\begin{equation}
a_{\mathrm{out}}(t)=a_{\mathrm{in}}(t)+\sqrt{\kappa }a(t).  \label{re}
\end{equation}%
Now assuming a large enough $\kappa$ to make sure  that we have a
weak excitation by the single-photon pulse on the atom initially
prepared in the ground state, i.e., keeping $\langle \sigma
_{z}\rangle =-1$ throughout our operation, we can adiabatically
eliminate the
cavity mode and arrive at the input-output relation of the cavity field,%
\begin{equation}
r(\omega _{\mathrm{p}})=\frac{[i(\omega _{c}-\omega _{\mathrm{p}})-\frac{%
\kappa }{2}][i(\omega _{0}-\omega _{\mathrm{p}})+\frac{\gamma }{2}]+g^{2}}{%
[i(\omega _{c}-\omega _{\mathrm{p}})+\frac{\kappa }{2}][i(\omega _{0}-\omega
_{\mathrm{p}})+\frac{\gamma }{2}]+g^{2}},  \label{rela}
\end{equation}%
where $r(\omega _{\mathrm{p}})\equiv \frac{a_{\mathrm{out}}(t)}{a_{\mathrm{in%
}}(t)}$ is the reflection coefficient for the atom-cavity system, and we
have assumed in Eq. (5) that the input field of the two-level atom, $b_{%
\mathrm{in}}(t)$, as a vacuum field, gives negligible contribution
to the output cavity field $a_{\mathrm{out}}(t)$. Eq. (\ref{rela})
is a general expression for various cases. For example, in the case
of $g=0$, Eq. (5) can recover the previous result for an empty
cavity \cite{Walls94},
\begin{equation}
r_{0}(\omega _{\mathrm{p}})=\frac{i(\omega _{c}-\omega _{\mathrm{p}})-\frac{%
\kappa }{2}}{i(\omega _{c}-\omega _{\mathrm{p}})+\frac{\kappa }{2}}.
\label{r0}
\end{equation}
Eq. (\ref{rela}) also fits very well the results in
\cite{Duan04,Feng}: If $g$ is dominant with respect to other
parameters, $r(\omega_{\mathrm{p}})$ would be 1, implying that the
input photon remains unchanged when it is output. Using the models
in \cite{Duan04,Feng}, we could explain the dominant $g$ case as
that, due to the strong resonant coupling between the cavity field
and the atom, the energy levels of the cavity will be shifted by the
large vacuum Rabi splitting, yielding a large detuning between the
dressed-cavity-mode and the single photon which is of the same
frequency as that of the original cavity.  Equivalently the total
system can be seen as a photonic pulse interacting with a
far-detuned bare cavity. As a result of Eq. (\ref{r0}) we have the
reflective coefficient being 1. This implies that the photon enters
and then leaks out of the cavity without being absorbed by the
cavity mode. In contrast, in the case that the cavity mode is
far-detuned with respect to the confined atom, no level shift would
occur in the cavity. Since the single photon is of the same
frequency as that of the cavity, we can equivalently see the model
as the photon interacting resonantly with a bare cavity, which
yields $r_{0}(\omega _{\mathrm{p}})=-1$ from Eq. (\ref{r0}).
From Eq. (5) the large detuning between
the cavity mode and the atom means $g=0$. As a result, we have
$r(\omega _{\mathrm{p}})=-1$. In this sense, we may argue that the
controlled phase flip designed in \cite{Duan04,Feng} seems a
special case of Eq. (\ref{rela}).

\begin{figure}[tbp]
\begin{center}
\scalebox{0.44}{\includegraphics{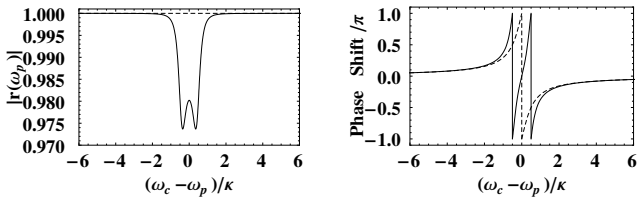}}
\end{center}
\caption{The absolute value and phase shift of the reflection coefficient $r(%
\protect\omega _{p})$ as functions of the detuning between the input pulse
and the cavity modes, with (solid line) and without (dashed line) the
presence of the atom, where $\protect\omega _{0}=\protect\omega _{c}$, $%
\protect\gamma /\kappa =0.01$, and $g/\protect\kappa =0.5$ ($g=0$)
for solid line (for dashed line).} \label{r}
\end{figure}

In fact, the key condition for Eq. (5) is $\langle \sigma
_{z}\rangle =-1$. So as long as $\kappa$ is large enough, which
makes $\langle \sigma _{z}\rangle =-1$ always satisfied, Eq. (5)
should work well even if $g$ is bigger than $\kappa$. As this
condition was met in \cite{Duan04,Feng}, it is not strange that we
could apply Eq. (5) to some physics in \cite{Duan04,Feng}. For more
general cases, we plot Fig. \ref{r} by Eq. (5) with the values from
\cite {Dayan08} for the absolute value and the phase shift of the
reflection coefficient. In the case of an empty cavity, the pulse
will experience a perfect reflection in the whole frequency regime,
i.e. $\left\vert r(\omega
_{\mathrm{p}})\right\vert =1$. The phase shift is $\pm\pi$ at $\omega _{%
\mathrm{p}}=\omega _{c}$, but reduces to zero rapidly with the frequency of
the input pulse deviating from the resonant point. In the case of an atom
presented in the cavity, the coupling between the cavity field and the atom
will shift the cavity mode, which leads to the vacuum splitting. Such a
splitting also makes the reflection coefficient a corresponding splitting,
as shown in the absolute value and the phase shift of the reflective
coefficient in Fig. 1. So it is the interaction between the cavity field and
the atom that induces weak absorption of the photon with detuned frequency
by the cavity mode and also decreases the reflection rate $\left\vert
r(\omega _{\mathrm{p}})\right\vert $. However, as shown in Fig. \ref{r},
even for a bad cavity with strong damping, the decrease of the reflection
rate is very small so that we can still have $\left\vert r(\omega _{\mathrm{p%
}})\right\vert \approx 1$.

\begin{figure}[tbp]
\begin{center}
\scalebox{1.0}{\includegraphics{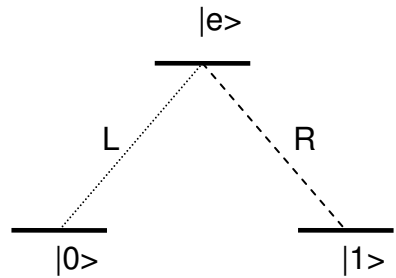}}
\end{center}
\caption{The relevant atomic structure subject to a bimodal cavity
field, where the lower levels are Zeeman sublevels of the ground
state and the upper level is the excited one. The dashed line
represents the $\text{R}$ circularly polarized mode and the dotted
line represents the $\text{L}$ circularly polarized mode.}
\label{level1}
\end{figure}

\section{Faraday rotation of the photonic polarization}

In the preceding section, we did not consider the polarization
degrees of freedom of the single photon. In this section, we show
that, with the general input-output relation, we can obtain a
rotation regarding the polarization of the single photon pulse after
the input-output process, known as the Faraday rotation
\cite{Julsgaard01}.

To explain the mechanism of Faraday rotation in cavity-QED, we may consider
the atom with the level structure as in Fig. \ref{level1}. The states $%
|0\rangle $ and $|1\rangle $ correspond to the Zeeman sublevels of
an alkali atom in the degenerate ground state, and $|e\rangle $ is
the excited state. We assume that the transitions of $|e\rangle
\leftrightarrow |0\rangle $ and $|e\rangle \leftrightarrow |1\rangle
$ are due to the coupling to two degenerate cavity modes
$a_{\mathrm{L}}$ and $a_{\mathrm{R}}$ with left (\textrm{L}) and
right (\textrm{R}) circular polarization, respectively. If the atom
is
initially prepared in $|0\rangle $,\ the only possible transition is $%
|0\rangle \rightarrow |e\rangle $, which implies that only the
\textrm{L} circularly polarized single-photon pulse
$|\mathrm{L}\rangle $ will take action. So
from Eq. (\ref{rela}) we have the output pulse related to the input one as $%
|\Psi _{\mathrm{out}}\rangle _{\mathrm{L}}=r(\omega _{\mathrm{p}})|\mathrm{L}%
\rangle \approx e^{i\phi }|\mathrm{L}\rangle $ with $\phi $ the
corresponding phase shift determined by the parameter values. It
also means that an input \textrm{R} circularly polarized
single-photon pulse $|\text{R}\rangle $ would only sense the empty
cavity. As a result, the corresponding output
governed by Eq. (\ref{r0}) is $|\Psi _{\mathrm{out}}\rangle _{\mathrm{R}%
}=r_{0}(\omega _{\mathrm{p}})|\mathrm{R}\rangle =e^{i\phi _{0}}|\mathrm{R}%
\rangle $ with $\phi_{0}$ a phase shift different from $\phi$. Therefore,
for an input linearly polarized photon pulse $|\Psi _{\mathrm{in}}\rangle =%
\frac{1}{\sqrt{2}}(|\mathrm{L}\rangle +|\mathrm{R}\rangle )$, the
output pulse is
\begin{equation}
|\Psi _{\mathrm{out}}\rangle _{-}=\frac{1}{\sqrt{2}}(e^{i\phi }|\mathrm{L}%
\rangle +e^{i\phi _{0}}|\mathrm{R}\rangle ).
\end{equation}%
The polarization degrees of freedom of a linearly polarized optical
field can be characterized by the Stokes vector
$\mathbf{S}=(S_{x},S_{y},S_{z})$ with
\cite{Hammerer04},%
\begin{eqnarray}
S_{x} &=&\frac{1}{2}(a_{\mathrm{L}}^{\dag }a_{\mathrm{R}}+a_{\mathrm{R}%
}^{\dag }a_{\mathrm{L}}),  \notag \\
S_{y} &=&\frac{1}{2i}(a_{\mathrm{L}}^{\dag }a_{\mathrm{R}}-a_{\mathrm{R}%
}^{\dag }a_{\mathrm{L}}),  \notag \\
S_{z} &=&\frac{1}{2}(a_{\mathrm{L}}^{\dag }a_{\mathrm{L}}-a_{\mathrm{R}%
}^{\dag }a_{\mathrm{R}}),
\end{eqnarray}
where $a_{k}$ ($a_{k}^{\dagger}$) with $k=\mathrm{L}$ or
$\mathrm{R}$ is the annihilation (creation) operator regarding
different polarization. It is
easily verified that $|\Psi _{\mathrm{in}}\rangle =\frac{1}{\sqrt{2}}(|%
\mathrm{L}\rangle +|\mathrm{R}\rangle )$ corresponds to $\mathbf{S}_{\mathrm{%
in}}=\frac{1}{2}(1,0,0)$ and Eq. (7) could be rewritten as $\mathbf{S}_{%
\mathrm{out}}=\frac{1}{2}(\cos (\phi _{0}-\phi ),\sin (\phi
_{0}-\phi ),0)$, for which we define $\Theta _{F}^{-}=\phi
_{0}-\phi$ to be Faraday rotation.

Similarly, if the atom is initially prepared in $|1\rangle $, then only the $%
\mathrm{R}$ circularly polarized photon could sense the atom,
whereas the $\mathrm{L}$ circularly polarized photon only interacts
with the empty cavity. So we have,
\begin{equation}
|\Psi _{\mathrm{out}}\rangle _{+}=\frac{1}{\sqrt{2}}(e^{i\phi _{0}}|\mathrm{L%
}\rangle +e^{i\phi }|\mathrm{R}\rangle ),
\end{equation}
where the Faraday rotation is $\Theta _{F}^{+}=\phi -\phi _{0}$.

\section{Entanglement generation of the atomic states by Faraday
rotation}

\begin{figure}[tbp]
\begin{center}
\scalebox{0.86}{\includegraphics{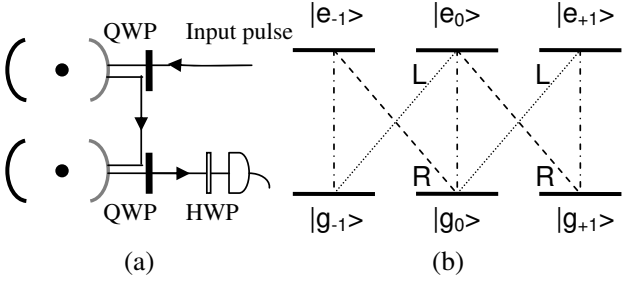}}
\end{center}
\caption{(a) Schematic for the generation of entangled atomic states
in fiber-connected cavities by Faraday rotation, where the atoms are
resonantly coupled to the cavities, respectively, and the input
photon pulse is detuned from the cavity modes. The bold lines
represent quarter-wave plates (QWPs), which achieve
$|\mathrm{L}\rangle\leftrightarrow |\mathrm{h}\rangle$ and
$|\mathrm{R}\rangle\leftrightarrow |\mathrm{v}\rangle$ of the
photon. HWP denotes a half-wave plate. (b) The relevant atomic
structure is subject to the bimodal cavity field, where the dashed line
represents the $\mathrm{R}$-polarized mode, the dotted line represents the $%
\mathrm{L}$-polarized mode and the dot-dashed line denotes the laser
pump pulse.} \label{level}
\end{figure}

Using the Faraday rotation introduced in last section, we could
generate entanglement between the atoms in spatially separate
cavities with low-Q factors, as plotted in
Fig. 3(a). We consider the level structure of each trapped atom as in Fig. %
\ref{level}(b), where $|g_{i}\rangle $ and $|e_{i}\rangle $ ($i=\pm 1,0$)
are the degenerate Zeeman sublevels of a typical alkali atom with $F=1$. As
the atom is in resonance with the cavity modes, the possible
cavity-mode-induced transitions are $|g_{-1}\rangle \leftrightarrow
|e_{0}\rangle $ and $|g_{0}\rangle \leftrightarrow |e_{+1}\rangle $ (or $%
|g_{+1}\rangle \leftrightarrow |e_{0}\rangle $ and $|g_{0}\rangle
\leftrightarrow |e_{-1}\rangle $) by absorbing or emitting a
\textrm{L-}(or \textrm{R-}) circularly polarized photon. The
transitions $|g_{i}\rangle \rightarrow |e_{i}\rangle $ ($i=\pm 1,0$)
can be realized by a classical laser pulse. To entangle two atoms
confined in spatially separate locations, we encode the
qubits in $|g_{-1}\rangle $ and $|g_{+1}\rangle $. As shown in Fig. \ref%
{level}(a), we input a single photon pulse in superposition of horizontal
and vertical polarizations, i.e., $|\Psi _{%
\mathrm{in}}\rangle =\frac{1}{\sqrt{2}}(|\mathrm{h}\rangle +|\mathrm{v}%
\rangle )$ (which changes to $|\Psi _{%
\mathrm{in}}\rangle =\frac{1}{\sqrt{2}}(|\mathrm{L}\rangle +|\mathrm{R}%
\rangle )$ after going through the quarter-wave plate (QWP)) into
the first cavity, and then we direct the output pulse by a fiber to
the second cavity. The detection of the output photon from the
second cavity, assisted by a QWP and a HWP, would yield the atoms to
be entangled.

Specifically if the two atoms are prepared in $|\psi _{n}\rangle =\alpha_{n}
|g_{-1}\rangle _{n}+\beta _{n}|g_{+1}\rangle _{n} = \alpha
_{n}|0\rangle _{n}+\beta _{n}|1\rangle _{n}$ with $n=1,2$ for different
cavities, the photonic input and output regarding the first cavity yield,
\begin{equation}
|\Psi _{\mathrm{in}}\rangle |\psi _{1}\rangle \rightarrow \alpha
_{1}|0\rangle _{1}|\Psi _{\mathrm{out}}\rangle _{-}+\beta _{1}|1\rangle
_{1}|\Psi _{\mathrm{out}}\rangle _{+},
\end{equation}%
which means an entanglement between the photonic and atomic qubits due to
different Faraday rotations. The photon going in and then reflected out of
the second cavity corresponds to,
\begin{eqnarray}
&&[\alpha _{1}|0\rangle _{1}|\Psi _{\mathrm{out}}\rangle _{-}+\beta
_{1}|1\rangle _{1}|\Psi _{\mathrm{out}}\rangle _{+}]|\psi _{2}\rangle  \notag
\\
&\rightarrow &\alpha _{1}\alpha _{2}\frac{1}{\sqrt{2}}(e^{i(\phi +\phi
^{\prime })}|\mathrm{h}\rangle +e^{i(\phi _{0}+\phi _{0}^{\prime })}|\mathrm{%
v}\rangle )|0\rangle _{1}|0\rangle _{2}  \notag \\
&&+\beta _{1}\beta _{2}\frac{1}{\sqrt{2}}(e^{i(\phi _{0}+\phi
_{0}^{\prime })}|\mathrm{h}\rangle +e^{i(\phi +\phi ^{\prime
})}|\mathrm{v}\rangle
)|1\rangle _{1}|1\rangle _{2}  \notag \\
&&+\alpha _{1}\beta _{2}\frac{1}{\sqrt{2}}(e^{i(\phi +\phi _{0}^{\prime })}|%
\mathrm{h}\rangle +e^{i(\phi _{0}+\phi ^{\prime
})}|\mathrm{v}\rangle
)|0\rangle _{1}|1\rangle _{2}  \notag \\
&&+\beta _{1}\alpha _{2}\frac{1}{\sqrt{2}}(e^{i(\phi _{0}+\phi ^{\prime })}|%
\mathrm{h}\rangle +e^{i(\phi +\phi _{0}^{\prime
})}|\mathrm{v}\rangle )|1\rangle _{1}|0\rangle _{2}, \notag \\
\label{ewe}
\end{eqnarray}%
where the actions of the QWPs have been included.
As we may adjust the frequency of the input pulse to $\omega _{\mathrm{p}%
}=\omega _{c}-\kappa /2$, we actually have $\phi =\phi ^{\prime }=\pi $ from
Eq. (\ref{rela}) with $g=\kappa /2$ and $\omega _{0}=\omega _{c}$, and we
have $\phi _{0}=\phi _{0}^{\prime }=\pi /2$ from Eq. (\ref{r0}) with $\omega
_{0}=\omega _{c}$. So the output state becomes%
\begin{eqnarray}
&&\frac{1}{\sqrt{2}}(|\mathrm{h}\rangle -|\mathrm{v}\rangle )(\alpha
_{1}\alpha _{2}|0\rangle _{1}|0\rangle _{2}-\beta _{1}\beta _{2}|1\rangle
_{1}|1\rangle _{2})  \notag \\
&&-i\frac{1}{\sqrt{2}}(|\mathrm{h}\rangle +|\mathrm{v}\rangle )(\alpha
_{1}\beta _{2}|0\rangle _{1}|1\rangle _{2}+\beta _{1}\alpha _{2}|1\rangle
_{1}|0\rangle _{2}).  \label{ef}
\end{eqnarray}%
After the output photon goes through a HWP, which makes $(|%
\mathrm{h}\rangle +|\mathrm{v}\rangle )/\sqrt{2}\rightarrow |\mathrm{h}\rangle $ and $%
(|\mathrm{h}\rangle -|\mathrm{v}\rangle )/\sqrt{2}\rightarrow
|\mathrm{v}\rangle $ \cite {explain2}, we may detect the photon in
the $|\mathrm{h}\rangle $ state, yielding a projection onto the
atomic state as
\begin{equation}
|\Phi \rangle _{12}=\frac{1}{N_{1}}[\alpha _{1}\beta _{2}|0\rangle
_{1}|1\rangle _{2}+\beta _{1}\alpha _{2}|1\rangle _{1}|0\rangle _{2}],
\label{e1}
\end{equation}%
where $N_{1}$ is the normalization constant. Alternatively, we may also
detect the photon in the $|\mathrm{v}\rangle $ state, yielding
\begin{equation}
|\Phi ^{\prime }\rangle _{12}=\frac{1}{N_{2}}[\alpha _{1}\alpha
_{2}|0\rangle _{1}|0\rangle _{2}-\beta _{1}\beta _{2}|1\rangle _{1}|1\rangle
_{2}],  \label{e2}
\end{equation}%
with $N_{2}$ the normalization constant. The atomic states we obtained in
Eqs. (\ref{e1}) and (\ref{e2}) are entangled with arbitrary amount of
entanglement determined by $\alpha _{1}$, $\alpha _{2}$, $\beta _{1}$ and $%
\beta _{2}$.

It is straightforward to extend above operations to the cases involving
three atoms. For example, if the output photon pulse from the second cavity
is directed to get in and then reflected out of the third cavity, we can
obtain the atomic state,
\begin{eqnarray}
&&|\Phi \rangle _{123}=\frac{1}{N_{1}^{\prime }}[\alpha _{1}\alpha
_{2}\alpha _{3}|0\rangle _{1}|0\rangle _{2}|0\rangle _{3}-\beta _{1}\beta
_{2}\alpha _{3}|1\rangle _{1}|1\rangle _{2}|0\rangle _{3}  \notag \\
&&~~~~~-\alpha _{1}\beta _{2}\beta _{3}|0\rangle _{1}|1\rangle _{2}|1\rangle
_{3}-\beta _{1}\alpha _{2}\beta _{3}|1\rangle _{1}|0\rangle _{2}|1\rangle
_{3}],
\end{eqnarray}%
corresponding to the output state of the photon with polarization $(|\mathrm{%
h}\rangle +i|\mathrm{v}\rangle )$, and
\begin{eqnarray}
&&|\Phi ^{\prime }\rangle _{123}=\frac{1}{N_{2}^{\prime }}[\beta _{1}\beta
_{2}\beta _{3}|1\rangle _{1}|1\rangle _{2}|1\rangle _{3}-\alpha _{1}\alpha
_{2}\beta _{3}|0\rangle _{1}|0\rangle _{2}|1\rangle _{3}  \notag \\
&&~~~~~-\beta _{1}\alpha _{2}\alpha _{3}|1\rangle _{1}|0\rangle
_{2}|0\rangle _{3}-\alpha _{1}\beta _{2}\alpha _{3}|0\rangle _{1}|1\rangle
_{2}|0\rangle _{3}],
\end{eqnarray}%
with the output state as $(|\mathrm{h}\rangle -i|\mathrm{v}\rangle )$. By
using another HWP (with a different tilted angle from the previously used one)
to achieve $(|\mathrm{h}\rangle +i|\mathrm{v}\rangle )/\sqrt{2}\rightarrow |%
\mathrm{h}\rangle $ and $(|\mathrm{h}\rangle -i|\mathrm{v}\rangle
)/\sqrt{2}\rightarrow |\mathrm{v}\rangle $, we distinguish the
states of the output pulse by the single-photon detector. Therefore,
we can generate the three-qubit entangled states in Eqs. (15) and
(16) at our will. Evidently, the scheme can be generalized to the
situation involving more atoms.

\section{Application of the generated entanglement}

In this section, we carry out some quantum information processing
tasks using the generated atomic entanglement. The key integrant is the efficient
conversion between the quantum nodes and the flying qubits. We will first show a
conversion from the atomic entanglement to the photonic entanglement
by some single qubit rotations on the atoms. Then we transfer any
unknown photonic state to the atom trapped in a distant cavity and
transfer an unknown state from one atom to another.

\subsection{Entanglement conversion from atoms to photons}

Supposing that the atoms in two separate cavities have been entangled as in
Eq. (\ref{e1}) with $\alpha=\alpha_{1}\beta_{2}/N_{1}$ and $\beta=\beta_{1}%
\alpha_{2}/N_{1}$, we apply two $\pi $-polarized classical laser
pulses simultaneously on the two atoms, and the states of the atoms
change as,
\begin{equation}
\alpha |0\rangle _{1}|1\rangle _{2}+\beta |1\rangle _{1}|0\rangle
_{2}\rightarrow \alpha |e_{-1}\rangle _{1}|e_{+1}\rangle _{2}+\beta
|e_{+1}\rangle _{1}|e_{-1}\rangle _{2}.  \label{efp}
\end{equation}%
Due to the atomic decay subject to the cavity modes, the de-excitation from $%
|e_{-1}\rangle $ (or $|e_{+1}\rangle $) to $|g_{0}\rangle $ produce a $%
\mathrm{R}$- (or $\mathrm{L}$-) circularly polarized photon. Then
Eq. (\ref{efp}) becomes
\begin{eqnarray}
&&\alpha |e_{-1}\rangle _{1}|e_{+1}\rangle _{2}+\beta |e_{+1}\rangle
_{1}|e_{-1}\rangle _{2}  \notag \\
&\rightarrow &|g_{0}\rangle _{1}|g_{0}\rangle _{2}(\alpha
|\mathrm{R}\rangle _{1}|\mathrm{L}\rangle _{2}+\beta
|\mathrm{L}\rangle _{1}|\mathrm{R}\rangle _{2}),
\end{eqnarray}%
where $|\mathrm{L}\rangle _{k}$ (or $|\mathrm{R}\rangle _{k}$) is
the generated photonic state with $\mathrm{L}$ (or $\mathrm{R}$)
polarization from the $k$th cavity, and the photon will change the
polarization to $|\mathrm{h}\rangle _{k}$ (or $|\mathrm{v}\rangle
_{k}$) after going through the QWP. In Eqs. (17) and (18), we have
omitted the photonic states in vacuum for simplicity. The two
equations above clearly show that the entanglement is converted from
the atomic states to the states of the emitting photons, which is
actually a source of entangled photons and will be used for further
quantum information processing mission.

\subsection{State transfer from photonic qubit to atomic qubit via
entanglement}

Our another scheme is to use the atomic entanglement generated above for
state transfer from a photon to an atom in distance. Suppose that the atoms
have been prepared in a maximally entangled state $\frac{1}{\sqrt{2}}%
(|0\rangle _{1}|1\rangle _{2}+|1\rangle _{1}|0\rangle _{2})$. As we want to
transfer an unknown photonic state $(x|\mathrm{h}\rangle +y|\mathrm{v}%
\rangle )$ ($|x|^2+|y|^2=1$) via the atomic entanglement from one
side to another, the photon only needs to go through one of the
cavities. First we input the single photon to the cavity in its
local side (for convenience, we call it as first cavity and the
other one the second cavity), the state of the total system
changes as,
\begin{eqnarray}
&&\frac{1}{\sqrt{2}}(x|\mathrm{h}\rangle +y|\mathrm{v}\rangle )(|0\rangle
_{1}|1\rangle _{2}+|1\rangle _{1}|0\rangle _{2})  \notag \\
&\rightarrow &\frac{1}{\sqrt{2}}(-x|\mathrm{h}\rangle |0\rangle
_{1}|1\rangle _{2}+ix|\mathrm{h}\rangle |1\rangle _{1}|0\rangle _{2}  \notag
\\
&&+iy|\mathrm{v}\rangle |0\rangle _{1}|1\rangle _{2}-y|\mathrm{v}\rangle
|1\rangle _{1}|0\rangle _{2}),
\end{eqnarray}%
where the input-output related Faraday rotations and the action of
QWPs have been considered. Then
the output pulse from the first cavity goes through a HWP, which makes $(|%
\mathrm{h}\rangle +|\mathrm{v}\rangle )/\sqrt{2}\rightarrow |\mathrm{h}\rangle $ and $%
(|\mathrm{h}\rangle -|\mathrm{v}\rangle )/\sqrt{2}\rightarrow
|\mathrm{v}\rangle $, followed by a detection. Besides, to recover
the photonic state in the second atom, we have to make Hadamard
operation and a $\sigma _{z}$ measurement on the first atom.
Depending on different measurement results regarding the photon and
the first atom, we perform different single-qubit operations $M_{i}$
on the second atom, i.e.,
\begin{eqnarray}
\mathrm{h};+ &:&\text{ }M_{1}=-ie^{-i\frac{\pi }{4}\sigma _{x}},  \notag \\
\mathrm{h};- &:&\text{ }M_{2}=-i\sigma _{y}e^{-i\frac{\pi }{4}\sigma _{x}},
\notag \\
\mathrm{v};+ &:&\text{ }M_{3}=-i\sigma _{y}e^{i\frac{\pi }{4}\sigma _{x}},
\notag \\
\mathrm{v};- &:&\text{ }M_{4}=ie^{i\frac{\pi }{4}\sigma _{x}},
\end{eqnarray}%
and the photonic state on the second atom is recovered as $|\psi
_{f}\rangle =\frac{1}{\sqrt{2}}(x|0\rangle +y|1\rangle )$.

\subsection{State transfer from one atomic qubit to another via
entanglement}

With similar steps in above subsection, we can also transfer
states between two separate atoms. Starting from Eq. (\ref{ef}) with
$\alpha _{2}=\beta _{2}=1/\sqrt{2}$, we transfer an arbitrary
state ($\alpha _{1}|0\rangle_{1}+\beta _{1}|1\rangle _{1}$) of the
first atom to the second. To this end, we perform a Hadamard gate on
the first atom in
Eq. (\ref{ef}), which yields%
\begin{eqnarray}
&&\frac{1}{2\sqrt{2}}(|\mathrm{h}\rangle -|\mathrm{v}\rangle )[|0\rangle
_{1}(\alpha _{1}|0\rangle _{2}-\beta _{1}|1\rangle _{2})  \notag \\
&&+|1\rangle _{1}(\alpha _{1}|0\rangle _{2}+\beta _{1}|1\rangle _{2})]
\notag \\
&&-i\frac{1}{2\sqrt{2}}(|\mathrm{h}\rangle +|\mathrm{v}\rangle )[|0\rangle
_{1}(\beta _{1}|0\rangle _{2}+\alpha _{1}|1\rangle _{2})  \notag \\
&&+|1\rangle _{1}(-\beta _{1}|0\rangle _{2}+\alpha _{1}|1\rangle _{2})].
\end{eqnarray}%
By detecting the photon polarization and the states of the first
atom, the second atom would be projected onto four corresponding
states with equal probability. After the local operations
conditioned on the measurement results, the
unknown state, i.e. ($\alpha _{1}|0\rangle _{2}+\beta _{1}|1\rangle
_{2}$), on the second atom is reconstructed.

\section{discussion and summary}

Eq. (5) is the key result of the present paper, based on which we have made the schemes
for entanglement generation and state transfer. It is easy to check that Eq. (5) fits very well the
numerical results in \cite{Duan04}. As an analytical expression working for a wide range
of parametric values,  Eq. (5) should be very useful in the study of cavity QED.

The entanglement generation and state transfer in Sections IV and V
are sketched for the low-Q cavities, such as the achieved MTR
\cite{Dayan08} or the single-sided cavity \cite{kimble95}. One point
we have to mention is that we have not yet figured out the
controlled logic gate between the two distant atomic qubits based on
the Faraday rotation, although we could achieve entanglement in
between. So we could not achieve teleportation between the two atoms
from the conventional viewpoint \cite{explain}. Anyway we have shown
the possibility to transfer an atomic or a photonic qubit state to a
distant atomic qubit. Different from the standard teleportation
steps, we need local operations on the qubits on both sides as
well as classical information to achieve the transfer of a quantum
state.

The entanglement of two distant atomic qubits by a single photon in our
scheme is more efficient than that by interference of two photons emitted
from two respective atoms \cite{Duan03}. In the latter case, the
entanglement between the two atomic qubits relies on the two leaking photons
reaching at the beam splitter simultaneously, as well as the high efficiency
of the two detectors, which strongly restricts the success rate in real
implementation. In contrast, as only a single photon is involved in our
scheme, there is no requirement for simultaneous detection. So our scheme
should be more efficient than those using two detectors in the case of the
large inefficiency of current single-photon detector.

The imperfection in our scheme is the photon loss, which is
also a problem in previously published schemes with photon
interference. The photon loss occurs due to the cavity mirror
absorption and scattering, the fiber absorption, and the
inefficiency of the detector. As the successful detection of the
photon ensures the accomplishment of our implementation, the
photon loss actually only affects the efficiency of the scheme, but
not the fidelity of the entanglement generation. Moreover, even if
we implement the scheme with low success rate, due to the highly
efficient single-photon source, such as 10,000 single photons per
second \cite{singlephoton}, we are able to accomplish our schemes
within a short time. We may simply assess the implementation of our
schemes, where the failure rate due to atomic decay is about
$\gamma/g=2\%$. Moreover, the current dark count rate of the
single-photon detector is about 100 Hz, which can reduce the
efficiency of our scheme by a factor of $10^{-4}$. Other
imperfection rate, regarding the photon absorption by the fiber and
the scattering of the cavity mirror, can be assumed to
be 6$\%$. Thus the success rate of our implementation should be $%
(1-2\%)^{2}\times 10^{-4}\times(1-6\%)=0.009\%$, where the square is
due to two atoms involved. By using the generation rate $10^{4}$
$s^{-1}$ of single photons, the two atoms can be entangled
within 2 sec, provided that the two atoms are not very distant,
i.e., without considering the time of the photon traveling in
between. Evidently, our scheme is in principle scalable with the
single photon going through the spatially separate cavities one by
one. However, with more atoms (confined in cavities) involved, the
photon loss would be more serious and thereby it will take  a
longer time for a successful event.

Before ending the paper, we would like to reiterate the difference
of our present work from the relevant work published previously
\cite {Duan04, Feng, goto}. The previous schemes prefer to work in
strong coupling condition, i.e., $g\gg \kappa, \gamma$. The fidelity
of the operations decreases significantly with the decrease of
$g/\kappa$. In contrast, our study focuses on the input-output
process of the low-Q cavity. We have not only analytically presented
an important expression for the photonic reflection, which
straightforwardly leads to our understanding of Faraday rotation,
but also demonstrated a way to entanglement generation and state
transfer in the case of low-Q cavity. We argue the impossibility of
controlled -NOT gating in the case under our consideration, which is
different from the strong-coupling or other cases considered in
\cite{Duan04, Feng, goto}. Meanwhile, we have shown the preference
with a single photon to carry out QIP tasks.

In summary, the general reflection rate of input-output process
we have analytically presented can be applied to different cases
for cavity QED: with and without atoms confined, and with strong or
weak coupling, and particularly useful for weak coupling case. Based
on this general reflection rate and a recently achieved MTR, we have
proposed a scheme using cavities with low-Q factors to entangle
distant atoms by a single photon, to generate entangled photons, and
to transfer quantum state to a distant qubit. We argue that our work
would be useful for QIP in cavity QED with current technology.

\section*{Acknowledgements}

The work is supported by NUS Research Grant No. R-144-000-189-305.
J.H.A. also thanks the financial supports of the NNSF of China under
Grant No. 10604025 and the Fundamental Research Fund for Physics and
Mathematics of Lanzhou University under Grant No. Lzu05-02. M.F. is
partially supported by NNSF of China under Grant No. 10774163.

Note added: After completing the manuscript, we become aware of a
recent work \cite{hu} with some similarities for quantum dots in
micropillar cavities. But our study on the cavity-atom system,
strongly relevant to the latest progress of cavity QED experiments,
is feasible with current technology.


\begin{thebibliography}{99}
\bibitem{explain2} This unitary rotation can be achieved by a half-wave
plate tilted by 22.5$^{o}$ with respect to the horizontal
direction. With different tilted angles, different
unitary operations are realized on the photonic polarization.
\end{thebibliography}
\end{document}